\documentclass[conference]{IEEEtran}
\ifCLASSINFOpdf
\else
\fi

\usepackage{cite}
\usepackage{amsmath,amssymb,amsfonts}
\usepackage{algorithm}
\usepackage[noend]{algpseudocode}
\usepackage{multirow}
\usepackage{graphicx}
\usepackage{textcomp}
\usepackage{xcolor}
\usepackage{fancyvrb, fancybox}
\usepackage{cprotect}
\usepackage{tcolorbox}
\usepackage{framed}
\usepackage{caption}
\usepackage{graphicx}
\usepackage{subcaption}
\usepackage{float}
\usepackage{hyperref}

\begin{document}

\newcommand{\hongce}[1]{\textcolor{red}{[\textbf{hongce}: #1]}}
\newcommand{\cy}[1]{\textcolor{blue}{[\textbf{changyuan}: #1]}}
\newcommand{\fixme}[1]{\footnote{\textbf{\color{red}{FIXME:}} #1}}
\newcommand{\todo}[1]{\footnote{\textbf{\color{red}{TODO:}} #1}}
\newcommand{\todohere}[1]{{\textbf{\color{red}{TODO:}} #1}}

\title{FRAIG-BMC: Functional Reduction to Speed Up Bounded Model Checking}

\author{
\IEEEauthorblockN{Changyuan Yu, Wenbin Che, Hongce Zhang}
\IEEEauthorblockA{
Hong Kong University of Science and Technology (Guangzhou)
}
}

\maketitle

\IEEEpeerreviewmaketitle

\begin{abstract}
Bounded model checking (BMC) is a widely used technique for formal property verification (FPV), where the transition relation is repeatedly unrolled to increasing depths and encoded into Boolean satisfiability (SAT) queries. As the bound grows deeper, these SAT queries typically become more difficult to solve, posing scalability challenges. Howevefor, many FPV problems involve multiple copies of related circuits, creating opportunities to simplify the unrolled transition relation. Motivated by the functionally reduced and-inverter-graph (FRAIG) technique, we propose FRAIG-BMC, which incrementally identifies and merges functionally equivalent nodes during the unrolling process. By reducing redundancy, FRAIG-BMC improves the efficiency of SAT solving and accelerates property checking. Experiments  demonstrate that FRAIG-BMC significantly speeds up BMC across a range of applications, including sequential equivalence checking, partial retention register detection, and information flow checking.

\end{abstract}
\section{Introduction}
Hardware formal verification is a crucial process in the design and validation of electronic systems. It  mathematically proves that a hardware design adheres to its specification and does not exhibit unexpected behaviors or bugs.


Bounded model checking (BMC)~\cite{bmc} is a widely adopted technique in formal property verification (FPV), where the transition relation of a system is repeatedly unrolled up to a given depth, and the accumulated transition logic is encoded as a Boolean satisfiability (SAT) query in conjunctive normal form (CNF).
Typically, BMC constructs a monolithic SAT query for each bound to check if a safety property is violated. While BMC is effective in finding short counterexamples, 
its performance can degrade significantly as the bound increases, due to the  growth of the CNF encoding and correspondingly the difficulty in SAT solving.

In many FPV applications, such as sequential equivalence checking, the transition system is composed of multiple copies of similar modules. This composition creates opportunities for simplification that the standard BMC algorithm does not exploit.

Inspired by the work on the functionally reduced and-inverter-graph (FRAIG)~\cite{fraigs}, this paper proposes FRAIG-BMC, a technique that improves BMC performance by reducing functional redundancy within the unrolled AIG network. Our method incrementally identifies and merges functionally equivalent nodes across time frames.
Before constructing the monolithic CNF formula and query the SAT solver for property violation of each bound, 
it first makes use of simulation to identify potentially equivalent nodes and then formally check their equivalence using a SAT solver. Other techniques such as structural hashing and trivial logic simplification are also employed to reduce the size of CNF formula.
Experiments show that the proposed methods can successfully speed up BMC in large cases and help it search into deeper bounds more quickly.

The rest of the paper is organized as follows: the next section presents the background of BMC and FRAIG. Section~\ref{algorithm} introduces the proposed algorithm in detail. In Section~\ref{experiment}, we present the result of experiments. Finally, Section~\ref{related-works} discusses the related works and Section~\ref{conclusion} concludes the paper.

\section{Preliminaries}
\label{pre}

\subsection{Bounded Model Checking}

In this paper, we focus on the bit-level hardware model checking problem where the state transition system is represented using an And-Inverter-Graph (AIG). For a state transition system with 
an initial state predicate $ \mathit{I}$ and a transition relation $\mathit{T}$, BMC tries to find the violation of a safety property $P$ within a certain bound.
It first tests for the existence of  an unsafe state within the initial state set by checking if the following Boolean formula is satisfiable:

\begin{equation}
    \mathrm{SAT}?\left(I\left(s_0\right) \wedge \neg P\left(s_0\right)\right)
\end{equation}

If this formula is unsatisfiable, there does not exist an initial state assignment that violates $P$, and BMC will start to unroll the transition relation.
For a bound $k$ (where $k\ge 1$), the state transition from $s_0$ to $s_k$ is:

\begin{equation}
    T^k(s_0, s_k) = \bigwedge_{0\le i<k} T(s_i, s_{i+1})
\end{equation}
And the following formula is used to check for counterexamples of bound $k$:
\begin{equation}
\label{boundcheck}
    \mathrm{SAT}?\left(I(s_0) \wedge T^k(s_0, s_k) \wedge \neg P(s_k)\right)
\end{equation}

BMC first tests for the case of $k=1$ and then incrementally check for greater $k$ if no counterexamples of smaller bound exist.
As for the implementation, BMC relies on the incremental solving capability of the underlying SAT solver to achieve better performance. Compared to solving  SAT queries of each bound individually, incremental solving allows the reuse of conflict clauses learned in the prior bounds and therefore, is usually more efficient.

\subsection{FPV with Multiple Copies of Similar Circuits}\label{sec::II-B}
In many FPV applications, the circuit under verification is composed of multiple copies of similar, or even functionally equivalent, sub-circuits. 
For example, in hardware side-channel detection, a shadow copy of the original circuit is instantiated to capture the implicit information flow, with both the original and shadow logic evaluated in parallel~\cite{ifc}. Similarly, in sequential equivalence checking, two functionally equivalent designs---often a specification and an optimized implementation---are instantiated side by side, and their outputs are compared across all time frames.
When running BMC on these problems, the unrolled transition relation will contain sub-formulas that can be unified and therefore, the final formula for SAT solving could be simplified.
 We argue that it is critical to 
 leverage the structural and functional similarity within the circuits of these applications for improving the scalability and efficiency of BMC.

\subsection{Functionally Reduced AIG (FRAIG)}

Functionally reduced AIG (FRAIG)~\cite{fraigs} is one such method to identify functionally equivalent circuit nodes. It employs random simulation to find circuit nodes that are potentially equivalent. It then uses SAT solving to check for the functional equivalence. If two nodes are found to be equivalent, they are merged together to reduce the size of the AIG circuit; otherwise, the satisfying assignment can be extracted to refine the candidate equivalent classes. The combination of random simulation and SAT solving is also referred to as SAT sweeping in the prior works~\cite{zhu2006sat,mishchenko2006improvements,satsweep-enhanced}.
 
 FRAIG can be applied 
 prior to the invocation of any model checking algorithms. 
 However, the standalone FRAIG procedure outside BMC cannot capture the opportunities of simplification introduced by the unrolling of state transition relations. 
 Therefore, in this paper, we propose to use an on-the-fly FRAIG procedure within the BMC algorithm to simplify the logic  resulted from unrolling $I(s_0) \wedge T^k(s_0, s_k)$.
 Compared to a standalone FRAIG procedure outside BMC, the integration could identify more equivalent circuit nodes due to unrolling.

\section{The algorithm of FRAIG-BMC }
\label{algorithm}


\subsection{Integration of FRAIG into BMC}
\begin{figure}[htbp]
    \centering
    \includegraphics[width=0.4\textwidth]{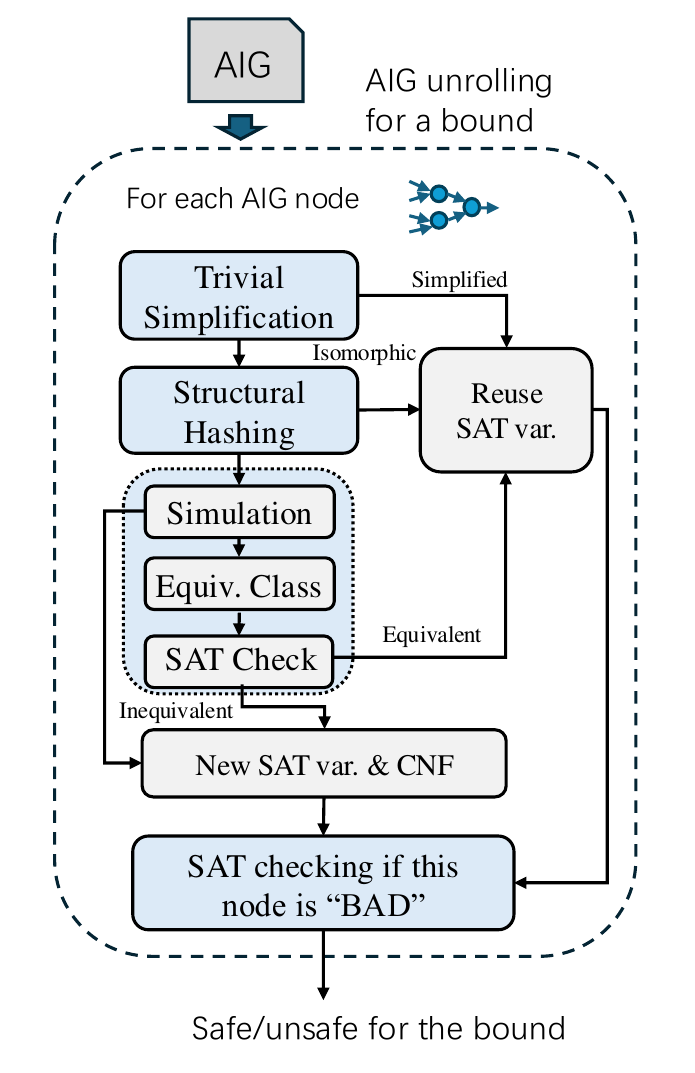}
    \caption{An illustration of integrating FRAIG in BMC unrolling}
    \label{fraig-bmc}
\end{figure}

In this section, we illustrate the detailed algorithm that integrates FRAIG into BMC. 
An overview of the FRAIG-BMC algorithm is shown in Fig.~\ref{fraig-bmc}. 
For a typical BMC algorithm,
at each bound, it unrolls the given transition relation by a topological traversal of the directed acyclic graph formed by the combinational logic of the circuit, which is represented as an AIG network.
For the first bound, initialized latch nodes are replaced by their initial value (either 0 or 1), while the uninitialized ones are treated in the same way as  input nodes, each of which is allocated with a SAT variable.
For all other bounds, latches will be replaced by the logic performing the state update in the previous bound.
For traditional BMC algorithms, each AND-gate corresponds to a new SAT variable, following the principle of Tseitin encoding.

As discussed in Section~\ref{sec::II-B}, in many FPV applications, there are structural and functional repetitions which can be unified to reduce the number of variables and clauses that BMC creates in unrolling. In FRAIG-BMC, we detect and reduce such repetitions following the three steps below:

\subsubsection{Trivial Logic Simplification}

For each AND-gate $n$ in AIG, we denote its two input nodes as $i_1$ and $i_2$.
\begin{itemize}
    \item If either $i_1$ or $i_2$ is constant false (0), then $n$ is resolved to false (0).
    \item If either $i_1$ or $i_2$ is constant true (1), then $n$ is resolved to the non-constant node or constant true (1) when both inputs are constant true (1). 
    \item If both inputs refer to the same nodes, then $n$ can be replaced by either of its input nodes.
\end{itemize}

Specifically, in the implementation, we keep a record of the mapping from the unrolled AIG nodes to the SAT variables. When an AND-gate is resolved to a constant or merge with another existing node, instead of creating a new SAT variable, the existing one will be reused.

\subsubsection{Structural Hashing}
For a node $n$ not simplified in the first step, we compute its structural hash based on its topology (the fanin node IDs and whether the inputs are complemented). This hash is used to detect whether an identical substructure has already been encountered. If a match is found with a previously registered node $p$, we conclude that $n$ is structurally redundant and merge it with $p$ by registering the SAT variable associated with $p$ for node $n$ in the node--variable map.

\subsubsection{Functional Redundancy Removal}
If node $n$ is structurally unique, we then check whether it is functionally equivalent to a previously encountered node. We compute a functional signature using random simulation and use this signature to group candidate equivalent nodes into an equivalence class (EC). If no other node is in this EC, then  $n$ is functionally unique  and we can encode it the same way as the traditional BMC by assigning a new SAT variable and adding the clauses in the SAT solver for this AND-gate.
If the EC contains other nodes, for each node $q$ in EC, we will query a SAT solver to check for the equivalence between $n$ and $q$.  If they are equivalent, node $n$ will be merged with node $q$, otherwise the satisfying assignment will be used to refine the simulation pattern.
In practice, we set a limitation to the size of the EC. When EC  is excessively large, it often suggests that the simulation may have a poor discriminating power. Instead of checking for equivalence with all nodes in EC, which could be time-consuming, we will treat it as if it is functionally unique and encode with new variables and clauses.

After the above three steps, only structurally and functionally unique nodes are assigned new SAT variables and encoded into CNF. All other nodes are merged, thereby avoiding redundant variables and clauses. Finally, for the node marked as ``bad'' (which is the negation of the safety property), we assert its corresponding SAT variable in the solver to check for property violation of that bound.  This incremental reduction process ensures that the CNF formula used to verify property $P$ grows more slowly, especially in the presence of repeated structures, significantly improving BMC solving efficiency.

\subsection{Handling Assumptions with Conditional Equivalence Checking}

FPV problems often contain assumptions (such as environmental constraints) that restrict the set of valid input sequences. Specifically, when verifying a property $P_k$ at bound $k$, the check must be performed under a set of constraints denoted as $C_k$, leading to the following  SAT formulation:

\begin{equation}
\label{constraint_bound_check}
    SAT?(I(s_0) \wedge T^k(s_0, s_k)\wedge C_k \wedge \neg P_k)
\end{equation}

These constraints affect not only property checking but also the process of detecting internal equivalence. In unconstrained scenarios, candidate equivalences are often discovered through random simulation patterns applied to input and uninitialized latch nodes. However, under constraints, many such random patterns may be invalid, as they violate $C_k$ and represent infeasible system behaviors. This can result in missed equivalences---nodes that behave identically under the constraint conditions may appear distinct in unconstrained simulation. To address this problem, we introduce two methods for constraint-aware equivalence detection, tailored to the strictness of the constraint set.

\subsubsection{Filtered Random Simulation}

When the constraints are relatively loose, a simple filtering approach is sufficient. 
We compute the cone of influence (COI) of the constraints, namely the set of nodes that are in the transitive fan-in of the  $C_k$ nodes.
After assigning random simulation values to the primary inputs and non-determined latches, we then simulate the constraint's COI using the random patterns and discard those that cause the constraints to evaluate to false. The remaining patterns, which satisfy all constraints, are retained and used for functional equivalence checking. This method maintains efficiency while enforcing correctness in scenarios with moderate constraint coverage.

\subsubsection{Constraint-Guided SAT Sampling}

For tighter constraints, filtered random simulation becomes ineffective, as most patterns are pruned, resulting in insufficient coverage for simulation-based equivalence detection. To overcome this limitation, we employ a SAT-based input pattern generation method.
For each bound $k$, we first extract the logic within constraint's COI and encode these logic with the constraint variable $C_k$ into a SAT query:
\begin{equation}
    SAT?(COI_k \wedge C_k)
\end{equation}

A satisfying assignment to this formula yields a simulation pattern that conforms to the constraints. To generate diverse patterns, we iteratively query the SAT solver with a blocking clause that negates the previously discovered solution and a random unit clause to increase the diversity of SAT solutions. The random unit clause can be removed if no satisfying solution is found.
Each new satisfying assignment provides a fresh constraint-compliant simulation pattern. Input nodes outside the COI of the constraints can still be assigned random values to ensure variability and increase simulation coverage.

\section{Experiments}
\label{experiment}


\begin{table*}[h]
\centering
\caption{Experiment results for SEC}
\label{retiming}
\begin{tabular}{c|ccc|ccc}
\hline
\multirow{ 2}{*}{Case Name}
 &
  \multicolumn{3}{c|}{\#. bound reached} &
  \multicolumn{3}{c}{\#. node merged} \\ \cline{2-7}
 &
  \multicolumn{1}{c|}{ABC-BMC3} &
  \multicolumn{1}{l|}{MiniSAT-AIGBMC} &
  \multicolumn{1}{l|}{FRAIG-BMC} &
  \multicolumn{1}{l|}{Trivial Logic Simp.} &
  \multicolumn{1}{l|}{Structural} &
  \multicolumn{1}{l}{Functional} 
  \\ \hline
ac97\_cra &
  \multicolumn{1}{c|}{13} &
  \multicolumn{1}{c|}{7} &
  \textbf{23} &
  \multicolumn{1}{c|}{283471} &
  \multicolumn{1}{c|}{224349} &
  42948 \\
aes &
  \multicolumn{1}{c|}{5} &
  \multicolumn{1}{c|}{5} &
  \textbf{39} &
  \multicolumn{1}{c|}{142632} &
  \multicolumn{1}{c|}{749580} &
  10331 \\
\multicolumn{1}{l|}{aes\_secworks} &
  \multicolumn{1}{c|}{1} &
  \multicolumn{1}{c|}{0} &
  \textbf{14} &
  \multicolumn{1}{c|}{210527} &
  \multicolumn{1}{c|}{512835} &
  42605 \\
des3\_area &
  \multicolumn{1}{c|}{1300} &
  \multicolumn{1}{c|}{\textbf{2648}} &
  1039 &
  \multicolumn{1}{c|}{597798} &
  \multicolumn{1}{c|}{1069415} &
  141296 \\
dft &
  \multicolumn{1}{c|}{\textbf{8}} &
  \multicolumn{1}{c|}{6} &
  3 &
  \multicolumn{1}{c|}{785821} &
  \multicolumn{1}{c|}{576932} &
  80503 \\
\multicolumn{1}{l|}{dynamic\_node} &
  \multicolumn{1}{c|}{7} &
  \multicolumn{1}{c|}{4} &
  \textbf{39} &
  \multicolumn{1}{c|}{501924} &
  \multicolumn{1}{c|}{606654} &
  86414 \\
fir &
  \multicolumn{1}{c|}{28} &
  \multicolumn{1}{c|}{17} &
  \textbf{67} &
  \multicolumn{1}{c|}{161292} &
  \multicolumn{1}{c|}{304375} &
  3235 \\
i2c &
  \multicolumn{1}{c|}{20} &
  \multicolumn{1}{c|}{17} &
  \textbf{243} &
  \multicolumn{1}{c|}{167597} &
  \multicolumn{1}{c|}{191316} &
  30433 \\
idft &
  \multicolumn{1}{c|}{\textbf{9}} &
  \multicolumn{1}{c|}{6} &
  3 &
  \multicolumn{1}{c|}{785837} &
  \multicolumn{1}{c|}{576952} &
  82056 \\
iir &
  \multicolumn{1}{c|}{13} &
  \multicolumn{1}{c|}{4} &
  \textbf{30} &
  \multicolumn{1}{c|}{108326} &
  \multicolumn{1}{c|}{203196} &
  8926 \\
sasc &
  \multicolumn{1}{c|}{39} &
  \multicolumn{1}{c|}{27} &
  \textbf{654} &
  \multicolumn{1}{c|}{420425} &
  \multicolumn{1}{c|}{333646} &
  47776 \\
sha256 &
  \multicolumn{1}{c|}{5} &
  \multicolumn{1}{c|}{2} &
  \textbf{36} &
  \multicolumn{1}{c|}{230723} &
  \multicolumn{1}{c|}{513771} &
  9877 \\
simple\_spi &
  \multicolumn{1}{c|}{38} &
  \multicolumn{1}{c|}{25} &
  \textbf{314} &
  \multicolumn{1}{c|}{223649} &
  \multicolumn{1}{c|}{213330} &
  40162 \\
ss\_pcm &
  \multicolumn{1}{c|}{39} &
  \multicolumn{1}{c|}{36} &
  \textbf{907} &
  \multicolumn{1}{c|}{444863} &
  \multicolumn{1}{c|}{286223} &
  45183 \\
tv80 &
  \multicolumn{1}{c|}{6} &
  \multicolumn{1}{c|}{6} &
  \textbf{41} &
  \multicolumn{1}{c|}{79615} &
  \multicolumn{1}{c|}{331415} &
  34782 \\
usb\_phy &
  \multicolumn{1}{c|}{42} &
  \multicolumn{1}{c|}{39} &
  \textbf{480} &
  \multicolumn{1}{c|}{265525} &
  \multicolumn{1}{c|}{159400} &
  25443 \\
wb\_dma &
  \multicolumn{1}{c|}{10} &
  \multicolumn{1}{c|}{8} &
  \textbf{84} &
  \multicolumn{1}{c|}{422158} &
  \multicolumn{1}{c|}{233668} &
  52646 \\ \hline
\end{tabular}
\end{table*}

\subsection{Benchmarks}
The FPV benchmarks used in our experiments originate from three application domains: (1) SEC: sequential equivalence checking of designs before and after retiming, (2) PartRet: detection and validation of partial retention registers for low-power designs~\cite{partret}, and (3) IFC: information flow checking on hardware modules, where SEC and IFC make use of circuits from OpenABC-D~\cite{openabc}. These problems contain dual copies of the similar circuits and could potentially benefit from on-the-fly simplification. All benchmark problems used in this study are publicly available at \url{https://github.com/ChangyuanYU/fraig-bmc_benchmark}. 


\subsection{Experiment Setup}

We implement FRAIG-BMC by extending the existing \texttt{aigbmc} tool from the AIGER library~\cite{aig}, incorporating the on-the-fly FRAIG procedure described earlier. The implementation uses the last released version of the state-of-the-art incremental SAT solver CadiCaL~\cite{cadical} as the underlying SAT solver for both property checking and functional equivalence verification. To ensure a fair comparison, we also adapt the baseline \texttt{aigbmc} tool to use the same CadiCaL backend, which we refer to as \texttt{CaDiCaL-AIGBMC}. Additionally, we include as a reference the BMC engine available in Berkeley-ABC~\cite{berkeleyabc}, invoked via the \texttt{bmc3} command, which represents the latest BMC implementation in that framework.
Experiments are conducted on a server with Intel Xeon 8375C processor and 256GB memory. We set the time limit to 3600 seconds.

\begin{figure}[t]
  \centering
  \includegraphics[width=0.8\linewidth]{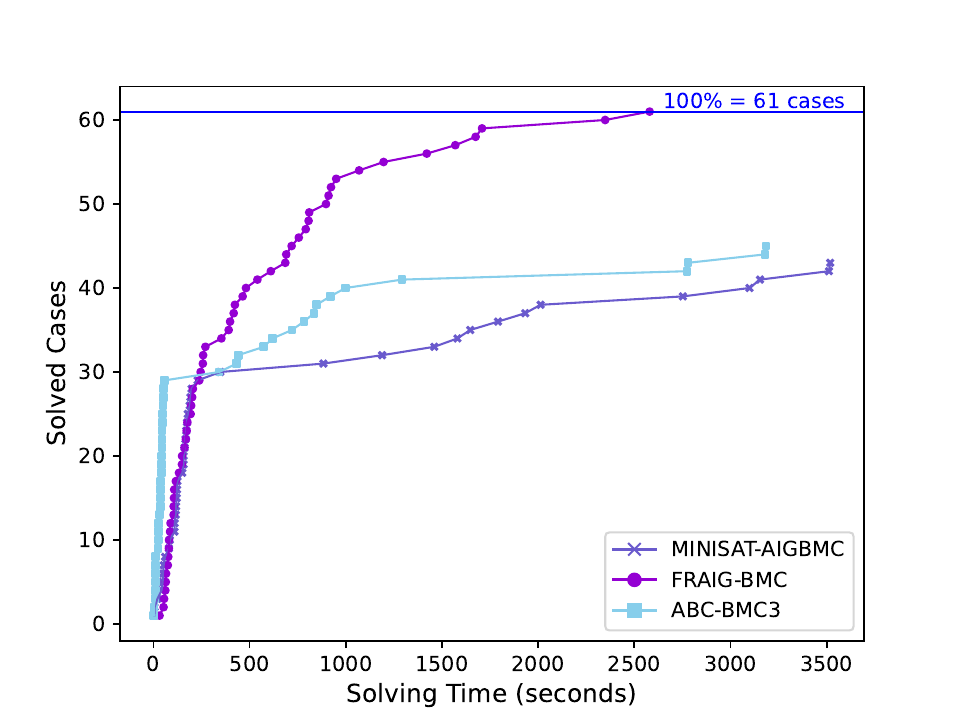}
  \caption{Number of solved cases vs. runtime for PartRet}
  \label{partret_compare}
  \vspace{-2mm}
\end{figure}

\begin{figure}[t]
  \centering
  \includegraphics[width=0.8\linewidth]{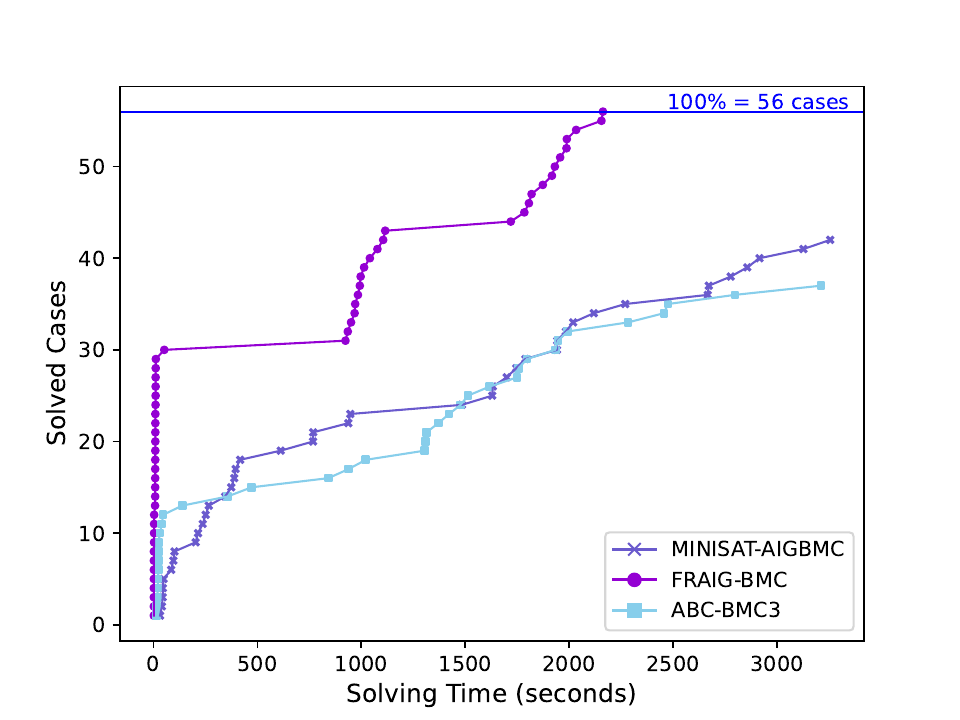}
  \caption{Number of solved cases vs. runtime for IFC}
  \label{ift_compare}
  \vspace{-5mm}
\end{figure}

\subsection{Experiment Results}
Results for the SEC problem set are shown in Table~\ref{retiming}. Note here that the input problems have previously gone through logic simplification in Berkeley-ABC to detect and remove functionally equivalent nodes before running model checking. However, FRAIG-BMC can still merge a significant number of nodes as shown by the last three columns.
This indicates that the standalone FRAIG procedure does not simplify the SAT queries to the same extend as an integrated FRAIG in BMC, as the prior typically does not factor in the sequential logic. Therefore, in most cases, FRAIG-BMC reaches far more bounds within the same time limit.
Fig.~\ref{partret_compare} and~\ref{ift_compare} depict the results of the experiments on the PartRet and IFC problem sets, which are expected to be unsafe.  FRAIG-BMC solves more cases compared to the traditional BMC engines thanks to its on-the-fly simplification.

\section{Related Works}
\label{related-works}

\noindent\textbf{BMC Acceleration.}
Prior works have explored a variety of approaches, such as parallel BMC~\cite{parallelBMC, tarmo} and partitioned solving strategies~\cite{cubeandconquer, partition}. While these techniques have achieved notable improvements in solving times, they are orthogonal to the contributions of this paper. In fact, both parallel and partitioned BMC instances can benefit from the proposed on-the-fly FRAIG reduction, particularly when the design under verification contains multiple related submodules.\looseness=-1

\vspace{0.5em} 
\noindent\textbf{FRAIG on CNF.}
Structural simplification and SAT sweeping have also been attempted at the CNF level by SAT solvers~\cite{satcomp21,satcomp22}, with techniques such as clausal congruence closure~\cite{clausal} and clausal equivalence sweeping~\cite{clausalequiv} integrated in Kissat.
However, because the original circuit structure is obscure in CNF, one has to recover useful structural information through reverse engineering, which takes notable efforts. Moreover, these CNF-level simplifications are typically applied only once to the static SAT instances and are not designed to support incremental SAT solving, which is crucial for an efficient BMC implementation.
This gap is addressed by the the on-the-fly FRAIG process proposed in this paper. \looseness=-1


\section{Conclusion}
\label{conclusion}

We present FRAIG-BMC, which integrates the SAT sweeping technique during BMC unrolling. Experiments show it is useful for problems involving multiple related hardware copies.\looseness=-1

\bibliographystyle{IEEEtran}
\bibliography{IEEEtran/ref}
\end{document}